\documentclass[12pt]{article}

\begin{document}
\title{Form factor of $\pi^0\gamma^*\gamma^*$ and contribution to muon g-2}
\author{Bing An Li\\
Department of Physics and Astronomy, University of Kentucky\\
Lexington, KY 40506, USA}

\maketitle
\begin{abstract}
An effective chiral theory of large $N_C$ QCD of mesons has been applied to
study the transition form factor of $\pi\gamma^*\gamma^*$. 
Besides the poles of vector mesons
an intrinsic form factor is found. The slope of the form factor is
predicted. The contribution of $\pi^0\gamma^*\gamma^*$ to muon g-2 is 
calculated. 
\end{abstract}
\newpage

A measurement of the muon g-2 with high accuracy has been reported[1] as
\[a_\mu=11659202(14)(6)\times10^{-10}.\]
The hadronic contributions to the muon g-2 consist of vacuum polarization, 
high order
corrections, and light-by-light scattering. The contribution of light-by-light 
scattering to the muon g-2($a^{lbl}$) has been studied[2,3]. The sign problem 
no longer exists.
The pseudoscalar poles play a dominant roles, especially, the pion pole.
Various transition form factor of $\pi^0\gamma^*\gamma^*$ have been
used in the calculation of $a^{lbl}$. The Wess-Zumino-Witten term, 
Vector Meson Dominance, and ENJL model are all used to obtain the transition
form factor of $\pi^0\gamma^*\gamma^*$. In Ref.[3] four different form factors
of $\pi^0\gamma^*\gamma^*$ are used to calculate the contribution to 
$a^{lbl}$.

The measurements of the form factor $\pi^0\gamma\gamma^*$ with one photon 
on mass 
shell  
have lasted for a long time[4,5].
In the timelike region the slope of the form factor of
$\pi^0\rightarrow\gamma e^+ e^-$
\[F(q^2)=1+a\frac{m^2_{e^+ e^-}}{m^2_{\pi^0}}\]
has been measured[5]. 
The measured value of the slope falls in a wide range of -0.24 to 0.12.
CELLO and CLEO[4] have measured the form factors of $P\gamma\gamma^*$ 
in the range of 
large $q^2$.
In CELLO's measurements the lower end of $q^2$ of 
$\pi^0\gamma\gamma^*$ 
is 0.5-0.8 $GeV^2$ and 1.5 $GeV^2$ from CLEO's measurements[4]. 
The slop of the form factor 
is obtained by
extrapolating the data to small values of $q^2$.
The PrimEx Coll. of JLab
is going to do direct precision measurements of the form factor of
$\pi^0\gamma\gamma^*$ at small values of $q^2$,
$0.001GeV^2 \leq q^2 \leq 0.5 GeV^2$[6].
On the other hand,
the form factor of $\pi^0\gamma\gamma^*$ has been studied by various
theoretical approaches[7].
 
We have proposed an effective chiral theory of large $N_C$ QCD of
pseudoscalar, vector, and axial-vector mesons[8].
In the limit $m_q\rightarrow 0$, this theory is chiral symmetric and
has dynamical chiral
symmetry breaking. The Lagrangian is expressed as
\begin{eqnarray}
{\cal L}=\bar{\psi}(x)(i\gamma\cdot\partial+\gamma\cdot v
+\gamma\cdot a\gamma_{5}+eQ\gamma\cdot A
-mu(x))\psi(x)-\bar{\psi(x)}M\psi(x)\nonumber \\
+{1\over 2}m^{2}_{0}(\rho^{\mu}_{i}\rho_{\mu i}+
\omega^{\mu}\omega_{\mu}
+a^{\mu}_{i}a_{\mu i}
+f^{\mu}f_{\mu})
\end{eqnarray}
where M is the quark mass matrix
\[\left(\begin{array}{c}
         m_{u}\hspace{0.5cm}0\\
         0\hspace{0.5cm}m_{d}
        \end{array}  \right ),\]
\(v_{\mu}=\tau_{i}\rho^{i}_{\mu}
+\omega_{\mu}\)
,
\(a_{\mu}=\tau_{i}a^{i}_{\mu}
+f_{\mu}\),
\(u=exp\{i\gamma_{5}(\tau_{i}\pi_{i}+
\eta)\}\), and m is the constituent quark mass which is related to dynamical 
chiral symmetry breaking.
The kinetic terms of mesons are generated by quark loops.
By integrating out the quark fields, the Lagrangian of mesons is
obtained.
In Eq.(1) the meson fields are needed to be normalized to
physical fields
\[\rho^i_\mu\rightarrow {1\over g}\rho^i_{\mu},\;\;\;
\omega_\mu\rightarrow {1\over g}\omega_{\mu},\]
where g is a universal coupling constant and defined as
\[g^2={1\over6}{F^2\over m^2},\]
and
under the cut-off 
\[{F^2\over16}=
\frac{m^2 N_C}{(2\pi)^4}\int\frac{d^4 k}{(k^2+m^2)^2}.\]

In order to cancel the mixing between $a^i_\mu$ and pion fields a
transformation
\[a^i_\mu\rightarrow {1\over g_{a}}a^i_\mu-{c\over g}\partial\pi^i \]
is introduced,
where
\[g_a={1\over g}(1-{1\over2\pi^2 g^2})^{-1}\;\;\;
c=\frac{f^2_\pi}{2gm^2_\rho}.\]
Physical pion field is defined as
\[\pi^i\rightarrow {2\over f_\pi}\pi^i,\]
where
\[f^2_\pi=F^2(1-{2c\over g}).\]

The universal coupling constant g and $f_\pi$ are two inputs.
g is determined
to be 0.39 by fitting
the decay rate of $\rho\rightarrow ee^+$ and
we take \(f_\pi=0.186GeV\).
$N_C$ expansion is revealed from this theory.
The tree diagrams are at leading order of $N_C$
expansion and loop diagrams of mesons are at higher orders. At low energies 
this theory goes back to the 
Chiral Perturbation Theory(ChPT) and the 10 coefficients of ChPT are 
determined[9].
The VMD is a natural result of this theory. 
Many physics processes have been calculated and theory agrees with data well
[11,12].  
Adler-Bell-Jackiw(ABJ) and
Wess-Zumino-Witten(WZW)[10] anomaly is the imaginary part of the Lagrangian 
of mesons.  	
The theory is phenomenological successful.

In this paper the study of the transition form factor 
of $\pi^0\gamma^*\gamma^*$ is presented.
It is interesting to mention the theoretical results of the form factor of 
charged pion[11,13].
The expressions of the VMD of $\rho$ and $\omega$ mesons 
up to fourth order in derivatives are derived from 
Eq.(1)
\begin{eqnarray}
\lefteqn{{1\over2}eg\{-{1\over2}F^{\mu\nu}(\partial_\mu \rho_\nu
-\partial_\nu \rho_\mu)+A^\mu j_\mu\}},\nonumber \\
&&{1\over6}eg\{-{1\over2}F^{\mu\nu}(\partial_\mu \omega_\nu
-\partial_\nu \omega_\mu)+A^\mu j^0_\mu\}.
\end{eqnarray}
Eqs.(2) are exactly the expressions of the VMD proposed by Sakurai[14]. 
According to Eqs.(2), there are two terms in the interactions between photon 
and mesons: photon couples to meson directly and photon via vector meson 
couples to mesons. By using the VMD, the form factor of charged pion is 
found(ignore $\rho-\omega$ mixing)[11,13]
\begin{eqnarray}
F_\pi (q^2) & = & f_{\rho \pi \pi }(q^2)
\frac{-m_\rho ^2+i\sqrt{q^2}\Gamma _\rho
(q^2)}{q^2-m_\rho ^2+i\sqrt{q^2}\Gamma _\rho (q^2)},
\end{eqnarray}
where $\Gamma _\rho(q^2)$ is the decay width of $\rho$[11].
$f_{\rho\pi\pi}(q^2)$ is the intrinsic form factor which is originated 
in the effects of quark loop[8]. To the fourth order in covariant derivative
\begin{equation}
f_{\rho\pi\pi}(q^2)=1+\frac{q^2}{2\pi^2 f^2_\pi}\{(1-{2c\over g})^2
-4\pi^2 c^2\}.
\end{equation}
The form factor of a charged pion consists of two parts: $\rho-$pole and 
intrinsic form factor. It is necessary to point out that  
$F_\pi(q^2)$ is derived from the VMD(2). 
The new factor in $F_\pi(q^2)$(3) is that the coupling between $\rho$($\gamma$)
and two pions is no longer a constant, instead, it is a function of $q^2$.
As pointed out in Ref.[15] that the $\rho-$ pole only form factor of 
pion decreases
too fast in space-like region and decreases too slow in the time-like region. 
The intrinsic form factor $f_{\rho\pi\pi}$ remedies these problems. The form
factor $F_\pi(q^2)$ agrees with data in both space-like and time-like 
regions($q^2
\sim 1.4GeV^2$)[11,13].  
On the other hand, the radius of charged pion
obtained from $\rho$ pole is
\[<r^2>_\pi=0.395 fm^2\]
and from Eq.(3) we obtain
\[<r^2>_\pi=0.452 fm^2.\]
The data is $(0.44\pm0.03)fm^2$[16]. $13\%$ of the radius comes from the intrinsic
form factor $f_{\rho\pi\pi}$.

In Ref.[13] we have studied the contribution of $e^+e^-\rightarrow\pi^+\pi^-$ 
to muon g-2.  
Besides the form factor of $\rho$ meson(3) a new mechanism of the 
$\rho-\omega$ mixing
and a new direct coupling $\omega\pi^+\pi^-$ are found from this 
effective chiral theory. Because of the new mechanism
the confrontation in the determination of the 
muon g-2 by using the data of $e^+ e^-\rightarrow\pi^+\pi^-$ and 
$\tau^-\rightarrow
\pi^0\pi^-\nu$ no longer exists. 

Now we use the same theory[8] to study
the transition form factor $\pi^0\gamma^*\gamma^*$. 
The two parameters, g and $f_\pi$, have been fixed, there is no new adjustable 
parameter in this study.
   
The $\pi^0\gamma^*\gamma^*$ is related to WZW anomaly. The vertices of the WZW 
anomaly can be derived from the imaginary part of the Lagrangian(1). 
The method deriving anomalous vertices is shown in Ref.[8].
Using the WZW anomalous vertices and VMD(2), the $\pi^0\gamma^*\gamma^*$ form
factor can be determined.
There are four processes shown in Fig.1, which contribute to the form factor.
In this paper we study the form factor $\pi^0\gamma^*\gamma^*$ at low energies.
The related vertices to the sixth order in derivatives are found from the
Lagrangian(1)
\begin{eqnarray}
\lefteqn{{\cal L}_{\pi^0\gamma\gamma}=-\frac{e^2}{4\pi^2 f_\pi}
\{1+{g^2\over2f^2_\pi}(1-{2c\over g})^2
(q^2_1+q^2_2+p^2)\}\pi^0\varepsilon^{\mu\nu\lambda\beta}\partial_\mu A_\nu
\partial_\lambda A_\beta,}\\
&&{\cal L}_{\pi^0\rho\gamma}=-\frac{e}{2g\pi^2 f_\pi}
\{1+\frac{g^2}{2f^2_\pi}(1-{2c\over g})^2 (q^2_1+q^2_2+p^2)\}
\pi^0\varepsilon^{\mu\nu\lambda\beta}\partial_\mu A_\nu
\partial_\lambda \rho_\beta,\\
&&{\cal L}_{\pi^0\omega\gamma}=-\frac{3e}{2g\pi^2 f_\pi}
\{1+\frac{g^2}{2f^2_\pi}(1-{2c\over g})^2(q^2_1+q^2_2+p^2)\}
\pi^0\varepsilon^{\mu\nu\lambda\beta}\partial_\mu A_\nu
\partial_\lambda \omega_\beta,\\
&&{\cal L}_{\rho\gamma}=-{e\over4}g(\partial_\mu
A_\nu-\partial_\nu A_\mu)
\{1-{1\over10\pi^2 g^2}{\partial^2\over m^2}\}(\partial^\mu \rho^{0\nu}
-\partial_\nu
\rho^{0\mu}),
\end{eqnarray}
where $q_1^2$, $q_2^2$, and $p^2$ are momentum of $\rho(\gamma)$, 
$\omega(\gamma)$, and $\pi^0$
respectively.
\begin{eqnarray}
\lefteqn{{\cal L}_{\omega\gamma}=-{e\over12}g(\partial_\mu A_\nu-\partial_\nu A_\mu)
\{1-{1\over10\pi^2 g^2}{\partial^2\over m^2}\}(\partial^\mu \omega^{\nu}
-\partial^\nu
\omega^{\mu})},\\
&&{\cal L}_{\pi^0\omega\rho}=-{3\over \pi^2 g^2 f_\pi}\pi^0
\varepsilon^{\mu\nu\lambda\beta}
\partial_\mu\rho^0_\nu\partial_\lambda\omega_\beta\{1+{1\over12m^2}
(1-{2c\over g})(
q^2_1+q^2_2+p^2)\},
\end{eqnarray}
where $q_1^2$, $q_2^2$, and $p^2$ are momentum of $\rho(\gamma)$, 
$\omega(\gamma)$, and $\pi^0$
respectively.
If the two photons and $\pi^0$ are on mass shell Eq.(5) goes back to ABJ anomaly
\begin{equation}
{\cal L}_{\pi^0 \rightarrow\gamma\gamma}=-\frac{\alpha}{\pi f_\pi}
\varepsilon^{\mu\nu\lambda\beta}\pi^0 \partial_\mu A_\nu \partial_\lambda
A_\beta.
\end{equation}

Using Eqs.(5-10), the amplitudes of the processes of Fig.1 are calculated 
in the
chiral limit
\begin{equation}
<\gamma_1\gamma_2|S|\pi^0>=-i(2\pi)^4\delta^4(p-q_1-q_2)\frac{1}
{\sqrt{8m_\pi\omega_1\omega_2}}
\varepsilon^{\mu\nu\lambda\beta}\epsilon_\mu(1)\epsilon_\nu(2)q_{1\lambda}q_{2\beta}
\frac{2\alpha}{\pi f_\pi}F(q^2_1,q^2_2),
\end{equation}
where
\begin{eqnarray}
F(q^2_1,q^2_2)& = &{1\over2}f_{\pi\rho\omega}
m^2_\rho m^2_\omega\{\frac{1}{(q^2_1-m^2_\rho)(q^2_2-m^2_\omega)}
+\frac{1}{(q^2_1-m^2_\omega)(q^2_2-m^2_\rho)}\},
\end{eqnarray}
in space-like region,
and $q^2_1$, $q^2_2$, and $p^2$ are momentum of two photons and pion 
respectively.
$f_{\pi\rho\omega}$ is the intrinsic form factor 
\begin{equation}
f_{\pi\rho\omega}=1+\frac{g^2}{2f^2_\pi}(1-{2c\over g})^2(q^2_1+q^2_2+p^2).
\end{equation}
$f_{\pi\rho\omega}$ is not a function of the momenta of two virtual photons
only,
is a function of the pion momentum too. This property of the intrinsic form 
factor $f_{\pi\rho\omega}$ makes the form factor $F(q^2_1, q^2_2)$(13)
different 
from any 
of the 
four form factors of Ref.[3].
It is the same as the form factor of charged pion, the transition form factor 
of $\pi^0\gamma^*\gamma^*$ consists of two parts: vector meson poles and an 
intrinsic
form factor which is calculated to $O(p^2)$. 

Put one photon on mass shell($\pi^0$ too), in the chiral limit 
the form factor of $\pi^0\gamma\gamma^*$ is 
derived from Eq.(13)
\begin{equation}
F_{\pi^0 \gamma\gamma^*}(q^2) = {1\over2}
\{1+\frac{g^2}{2f^2_\pi}(1-{2c\over g})^2 q^2\}
\{\frac{m^2_\rho}{m^2_\rho-q^2}+\frac{m^2_\omega}{m^2_\omega-q^2}\}.
\end{equation}
At very low energies we obtain
\begin{eqnarray}
F_{\pi\rho\omega}(q^2) & = & 1+a{q^2\over m^2_\pi},\\
&&a=
{m^2_\pi\over2}({1\over m^2_\rho}+{1\over m^2_\omega})
+{m^2_\pi\over2f^2_\pi}g^2(1-{2c\over g})^2.
\end{eqnarray}
\begin{equation}
a=0.0303+0.0157=0.046.
\end{equation}
The first number of Eq.(18) is from the poles of vector mesons and the second 
number is the contribution of 
the intrinsic form factor(14).
The contribution of the intrinsic form factor is $34\%$ of the value of a. 
The value of a is the prediction of this theory and is a test of the 
form factor(13).
 
Using the form factor(13) 
and the formulas presented in Ref.[3](Eqs.(3.4-3.6)),   
we calculate the lbl contribution of pion pole at low 
energies. As shown the form factor of $\pi^0\gamma^*\gamma^*$ is a 
function of $p^2$
(momentum of pion) too. To avoid confusion, the Eq.(3.4) of Ref.[3] is 
presented as
\begin{eqnarray}
\lefteqn{a^{lbl,\pi^0}=-\frac{e^6}{4\pi^4 f^2_\pi}\int \frac{d^4 q_1}{(2\pi)^4}\int
\frac{d^4 q_2}{(2\pi)^4}\frac{1}{q^2_1 q^2_2 (q_1+q_2)^2[(p+q_1)^2-m^2]
[(p-q_2)^2-m^2]}}\nonumber \\
&&\{F(q^2_1,(q_1+q_2)^2)|_{p^2=q^2_2}F(q^2_2,0)|_{p^2=q^2_2}
\frac{T_1(q_1,q_2;p)}{q^2_2-m^2_\pi}\nonumber \\
&&+F(q^2_1,q^2_2)|_{p^2=(q_1+q_2)^2}F((q_1+q_2)^2,0)|_{p^2=(q_1+q_2)^2}
\frac{T_2(q_1,q_2;p)}{(q_1+q_2)^2-m^2_\pi}\},
\end{eqnarray}
where $T_1$ and $T_2$ can be found from Eqs.(3.5 and 3.6) of Ref.[3].
Comparing with CELLO's data[4], we choose the upper limits of $q^2_1$ and 
$q^2_2$
(Euclid space) as $(0.5GeV)^2$.
The numerical result of Eq.(19) is
\[a^{lbl,\pi}=8.48\times10^{-10}.\]

To conclude, the form factor of $\pi^0\gamma^*\gamma^*$ at low energies is 
found.
Besides the vector meson poles an additional
intrinsic form factor caused by quark loop is predicted. The slope of the form
factor of $\pi^0\rightarrow\gamma\gamma^*$ is predicted. The contribution of
the intrinsic form factor is about $34\%$ of the slope. 
The contribution of the pion pole at low energies 
via lbl to the muon g-2 is calculated.  

This study is supported
by a DOE grant.

\pagebreak
\begin{flushleft}
{\bf Figure Captions}
\end{flushleft}
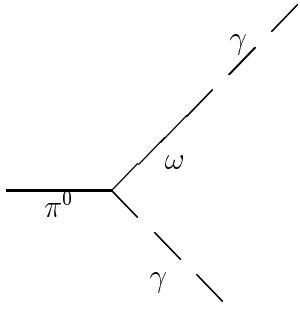
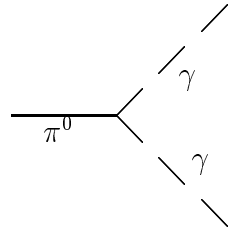
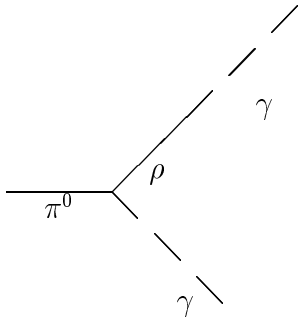
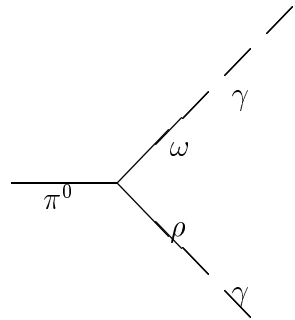
{\bf FIG. 1.} Processes contributing to $\pi^0\gamma^*\gamma^*$


\end{document}